\documentclass[twocolumn,twocolappendix]{aastex631}

\usepackage{natbib}
\usepackage{graphicx}	
\usepackage{amsmath}	
\usepackage{amssymb}	
\usepackage{bm}	
\usepackage{subfigure}
\usepackage{longtable}
\usepackage{array}
\usepackage{booktabs}
\usepackage{multirow}
\usepackage{xcolor}
\usepackage{lipsum}

\shorttitle{EP240414a: A LGRB jet in CSM}
\shortauthors{Hamidani, et al.}

\begin{document}
\title{EP240414a: A Gamma-Ray Burst Jet Weakened by an Extended Circumstellar Material}

\correspondingauthor{Hamid Hamidani}
\email{hhamidani@astr.tohoku.ac.jp}

\author[0000-0003-2866-4522]{Hamid Hamidani}
\affiliation{Astronomical Institute, Graduate School of Science, Tohoku University, Sendai 980-8578, Japan}

\author[0000-0003-2477-9146]{Yuri Sato}
\affiliation{Astronomical Institute, Graduate School of Science, Tohoku University, Sendai 980-8578, Japan}

\author[0000-0003-4299-8799]{Kazumi Kashiyama}
\affiliation{Astronomical Institute, Graduate School of Science, Tohoku University, Sendai 980-8578, Japan}

\author[0000-0001-8253-6850]{Masaomi Tanaka}
\affiliation{Astronomical Institute, Graduate School of Science, Tohoku University, Sendai 980-8578, Japan}

\author[0000-0002-3517-1956]{Kunihito Ioka}
\affiliation{Yukawa Institute for Theoretical Physics, Kyoto University, Kyoto 606-8502, Japan}

\author[0000-0003-2579-7266]{Shigeo S. Kimura}
\affiliation{Frontier Research Institute for Interdisciplinary Sciences, Tohoku University, Sendai 980-8578, Japan}
\affiliation{Astronomical Institute, Graduate School of Science, Tohoku University, Sendai 980-8578, Japan}

\begin{abstract}
The recent Einstein Probe (EP) event EP240414a exhibits several unusual observational features. 
Its prompt and afterglow emissions place it between long gamma-ray bursts (LGRBs) and low-luminosity GRBs (LLGRBs). 
The event is followed by a fast optical transient (AT~2024gsa), initially exhibiting a thermal-like spectrum but later evolving into an unusually red peak at $\sim 3-5$ days, which is difficult to explain with thermal emission. 
Using our generalized analytic framework for jet propagation in a circumstellar material (CSM; \citealt{2025arXiv250316242H}), we explore a scenario in which a conventional LGRB jet is launched in a progenitor surrounded by a dense CSM. 
For a CSM of $\sim 0.03 M_\odot$ extending to $\sim 3\times 10^{13}$ cm, we find that the jet is significantly weakened before breaking out, becoming ``barely failed'', an intermediate state between successful (LGRB) and completely failed (LLGRB) jets. 
This scenario naturally explains EP240414a's multi-wavelength observations, with the early thermal component produced by cocoon cooling emission, and the red peak explained by non-thermal afterglow emission from the mildly relativistic barely failed jet (and its inner-cocoon). 
Our work demonstrates the important role of extended CSM in shaping GRB jets and illustrates how early multi-wavelength follow-up observations can reveal the physically diverse nature of jet-driven transients.
\end{abstract}


\section{Introduction}
\label{sec:1}

The prompt emission of Gamma-Ray Bursts (GRBs) is explained by highly relativistic jets \citep{1973ApJ...182L..85K,1997ApJ...487L...1R,2004RvMP...76.1143P}. 
According to the collapsar model, long GRB (LGRB) jets are produced in the collapsing core of stripped-envelope massive stars, specifically Wolf-Rayet (WR) stars \citep{1993ApJ...405..273W,1999ApJ...524..262M}.
The association between LGRBs and broad-lined (BL) Type Ic supernovae (SNe; e.g. SN 1998bw; \citealt{1998Natur.395..670G}) confirms this model.

In contrast, low-luminosity GRBs (LLGRBs) exhibit lower isotropic energies ($E_{\rm iso} \sim 10^{48}-10^{50}$ erg) and softer spectra (e.g., GRB 060218 and GRB 100316D; \citealt{2006Natur.442.1008C,2006Natur.442.1014S,2011MNRAS.411.2792S}). 
LLGRBs are thought to be explained by shock breakout emission, implying that they do not require relativistic jets (\citealt{2007ApJ...667..351W,2012ApJ...747...88N}). 
In particular, it has been suggested that the weakness of prompt emission in LLGRBs (relative to LGRBs) can be explained by a dense circumstellar medium (CSM) around the progenitor, which prevents the jet from breaking out and produces a non-relativistic cocoon that powers shock breakout emission as it breaks out of the CSM (\citealt{2015ApJ...807..172N,2016MNRAS.460.1680I,2024arXiv241206736I}).

Recent observations provide growing evidence for extended CSM environments around stripped-envelope core-collapse SN (CCSNe; Types Ib, Ic, Ibn, and Icn) prior to explosion, 
and suggest that CSM masses range from $M_{\rm CSM} \sim 0.01-1 M_{\odot}$, extending out to radii of $R_{\rm CSM} \sim 10^{14}-10^{16}$ cm (e.g., \citealt{2007Natur.447..829P,2007ApJ...657L.105F,2008MNRAS.389..131P,2008ApJ...674L..85I,2006ApJ...653L.129B,2020A&A...643A..79S,2020MNRAS.492.2208C,2024ApJ...977..254D,2024ApJ...977....2P}; also see \citealt{2022ApJ...927...25M}). 

Studies of the interaction between LGRB jets and CSM suggest that the CSM can not only choke the jet, but also produce bright cooling emission from shocked CSM (i.e., cocoon), leading to transients in the form of fast blue optical transients (FBOTs; \citealt{2015ApJ...807..172N,2019ApJ...872...18M,2022ApJ...925..148S,2024PASJ...76..863S}; also see \citealt{2022MNRAS.513.3810G}; and \citealt{2020ApJ...900..193D}).

In a companion paper (\citealt{2025arXiv250316242H}), we developed a generalized analytic model that captures the dynamics of jet propagation in CSM environments.
Our model enables us to evaluate the outcome of a given LGRB jet propagating in a given CSM and predict the parameter space where successful/failed jets are produced and consequently where LGRBs/LLGRBs emission can emerge. 
Additionally, we identified an intermediate jet regime between completely failed and successful jets, which we refer to as ``barely failed" jets (previously discussed in \citealt{2022MNRAS.517..582E}).

The recent Einstein Probe (EP) mission, with its sensitivity to soft X-ray, has identified numerous X-ray transients, including EP240414a (\citealt{2024arXiv241002315S}). 
EP240414a exhibits an intermediate intensity between LLGRBs and LGRBs in both prompt and afterglow emissions (\citealt{2024arXiv241002315S,2025ApJ...982L..47V,2025ApJ...978L..21S,2024arXiv240919055B}). 
Here, assuming that certain WR stars possess a CSM and that EP240414a originates from a conventional collapsar event, we explore the necessary CSM conditions to explain EP240414a's unusual features. 
We show that with a CSM mass of $\sim 0.03 M_\odot$ extending to $\sim 3\times 10^{13}$ cm, the LGRB jet becomes barely failed, providing a scenario that coherently explains EP240414a's unique characteristics.

This paper is structured as follows.  
In Section \ref{sec:2}, we review the observational features of EP240414a.  
In Section \ref{sec:model}, we describe our proposed scenario.  
In Section \ref{sec:results}, we present our findings.  
In Section \ref{sec:diss}, we discuss their implications and provide concluding remarks.


\section{Observational properties}
\label{sec:2}
EP240414a is an X-ray transient detected by EP (\citealt{2024arXiv241002315S}) and subsequently observed across the X-ray, radio, optical and infrared (IR) bands.
In the optical and IR bands, a luminous ``FBOT" counterpart (AT~2024gsa, although red in color) was discovered, which evolved into a BL SN Ic at later times (SN~2024gsa, \citealt{2025ApJ...982L..47V,2025ApJ...978L..21S,2024arXiv240919055B}).
The event was located at $z=0.401$ ($D_{\rm L}\approx 2245$ Mpc; assuming a $\Lambda$CDM cosmology with a Hubble constant $H_0 = 67.7\,\mathrm{km\,s^{-1}\,Mpc^{-1}}$, $\Omega_{\mathrm{M}} = 0.309$, and $\Omega_{\Lambda} = 0.691$; \citealt{2016A&A...594A..13P}), at a large offset of $26$~kpc from the spiral galaxy J1246 (\citealt{2025ApJ...982L..47V,2024arXiv241002315S}). 
It is unclear whether it is physically offset or hosted by an unresolved faint-satellite galaxy.

Together, EP240414a, AT~2024gsa, and SN~2024gsa exhibit several unique features:
\begin{enumerate}
\renewcommand{\labelenumi}{\arabic{enumi})}
    \item Prompt emission: The isotropic emission energy is $E_{\rm iso} \sim 5 \times 10^{49}$ erg with a duration of $T_{90} \sim 155\,\mathrm{s}$. The emission is predominantly in the soft X-ray band ($E_\mathrm{peak} < 1$ keV). Both the isotropic energy and the peak energy are significantly lower than those of conventional LGRBs.
    Also, EP240414a stands as an outlier from the Amati-Yonetoku relations for LGRBs (see Figure 2 in \citealt{2024arXiv241002315S}).
    Furthermore, the observed properties of EP240414a do not align with the closure relationship between $E_{\rm iso}$, $T_{90}$, and $E_\mathrm{peak}$ predicted by the conventional shock breakout scenario [see equations (14)–(18) in \citealt{2012ApJ...747...88N}], which is typically valid for LLGRBs.
    \item Phase I ($\lesssim 1$ day): Early follow-up observations found a luminous FBOT (AT~2024gsa) consisting of two parts (phase I and phase II). 
    Spectroscopy in optical/IR during phase I (at $\sim 0.6$ day) shows a clear blue spectral slope ($F_\nu\propto \nu^{0.9}$) that is inconsistent with typical afterglow emission ($F_\nu\propto \nu^{-1}$; see Section 3.3 in \citealt{2025ApJ...982L..47V}). 
    \item Phase II ($\sim 1 - 10$ days): AT~2024gsa peaked with an absolute magnitude of $\sim -21$ mag (in r-band, at $\sim 4$ days). 
    At its peak, it exhibited a red color that challenges thermal emission scenarios (\citealt{2024arXiv241002315S,2025ApJ...982L..47V}). \cite{2025ApJ...978L..21S} pointed out that a thermal model would require a photospheric velocity of $\sim 0.7c$ to reproduce the color; which in turn would introduce additional issues due to relativistic effects on observer time and color shift.
    Spectroscopy in optical/IR around the peak (at $\sim 3-5$ days) shows a flat featureless spectrum, which is more compatible with afterglow scenarios (see Figure 2 in \citealt{2025ApJ...982L..47V}).
    \item Phase III ($\gtrsim10$ days): A BL Type Ic supernova, SN~2024gsa, was spectroscopically identified, exhibiting similarities to other LGRB-SNe such as SN~1998bw and SN~2006aj (\citealt{2024arXiv241002315S,2025ApJ...982L..47V}). However, SN~2024gsa appears significantly redder than typical BL Ic SNe (\citealt{2019A&A...621A..71T}; see Appendix \ref{sec:color}).
    \item Radio emission (days--months): 
    The emission is consistent with the lower luminosity end of LGRB afterglows, viewed on-axis, yet it remains orders of magnitude brighter than that of LLGRBs (\citealt{2024arXiv241002315S,2024arXiv240919055B}). 
    This rules out the scenario of a completely failed jet, and disfavors the presence of an energetic highly relativistic jet (\citealt{2024arXiv240919055B}).
\end{enumerate}

First, the observation in 4) points to an ordinary collapsar event involving a massive star (\citealt{1993ApJ...405..273W,1999ApJ...524..262M}).
From the properties in 1) and 5), we deduce that a completely choked/failed jet is ruled out.
Additionally, from 1) and 5), we deduce that a standard energetic LGRB jet is also disfavored.
From 2) and 3), we deduce that: 
although phase I of AT~2024gsa is thermal and could possibly be powered by the jet/SN CSM interaction (see \citealt{2025ApJ...982L..47V});
phase II of AT~2024gsa is inconsistent with thermal emission and could be afterglow powered (\citealt{2025ApJ...978L..21S}), which would be consistent with 5).

Based on these observations, we propose an intermediate scenario, where the jet is neither completely failed (as in LLGRBs) nor successful (as in LGRBs). 
We propose a ``barely failed" jet scenario to explain EP240414a/AT~2024gsa, where the jet is significantly weakened (energetically and relativistically) before breaking out (see \citealt{2022MNRAS.517..582E} and their Figure 6; also see \citealt{2025arXiv250316242H}).
We consider that this weakening is due to an extended CSM around the progenitor (similar but not identical to CSM proposed for LLGRBs in \citealt{2015ApJ...807..172N}; see \citealt{2025arXiv250316242H} for details on parameter space for this scenario) which also conveniently explains the early thermal emission ($\lesssim 1$ day; \citealt{2025ApJ...982L..47V}).

\section{The Physical Model}
\label{sec:model}
\subsection{The Model}
\label{sec:setup}

We consider a stripped massive star (WR; with a radius $r_0\sim 4\times 10^{10}$ cm) as the progenitor of a collapsar event (\citealt{1993ApJ...405..273W,1999ApJ...524..262M,2006ApJ...637..914W}).
We consider that the star is surrounded by an extended CSM, similar in nature (although not in scale) to those observed in stripped envelope SNe (see \citealt{2007Natur.447..829P,2007ApJ...657L.105F,2008MNRAS.389..131P,2008ApJ...674L..85I,2006ApJ...653L.129B,2020A&A...643A..79S,2020MNRAS.492.2208C,2024ApJ...977..254D,2024ApJ...977....2P}) which is assumed to be produced as the result of an intense mass-loss episode prior to the collapsar event.

\begin{figure*}
    \centering
    \includegraphics[width=0.49\linewidth]{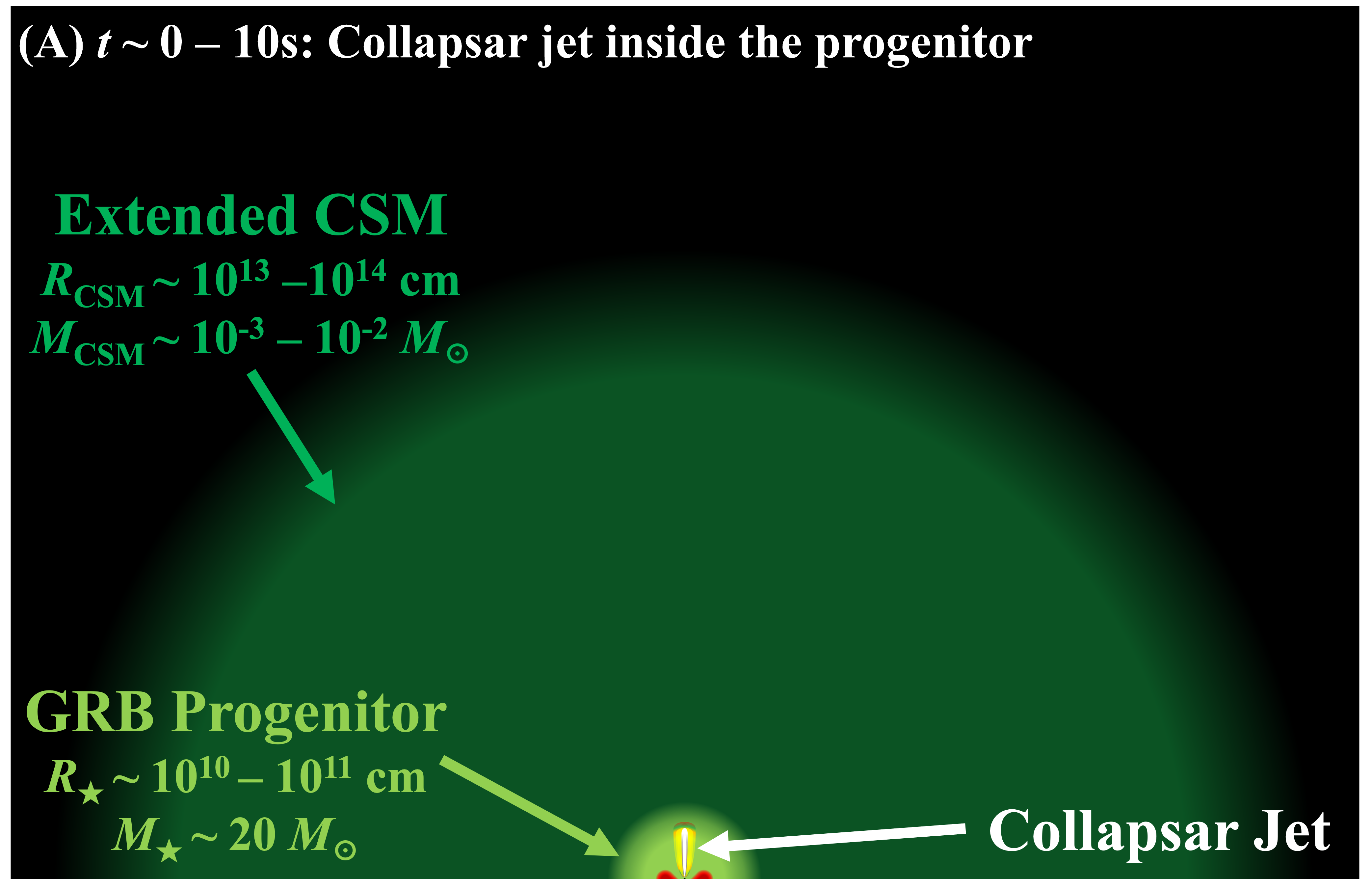} 
      \includegraphics[width=0.49\linewidth]{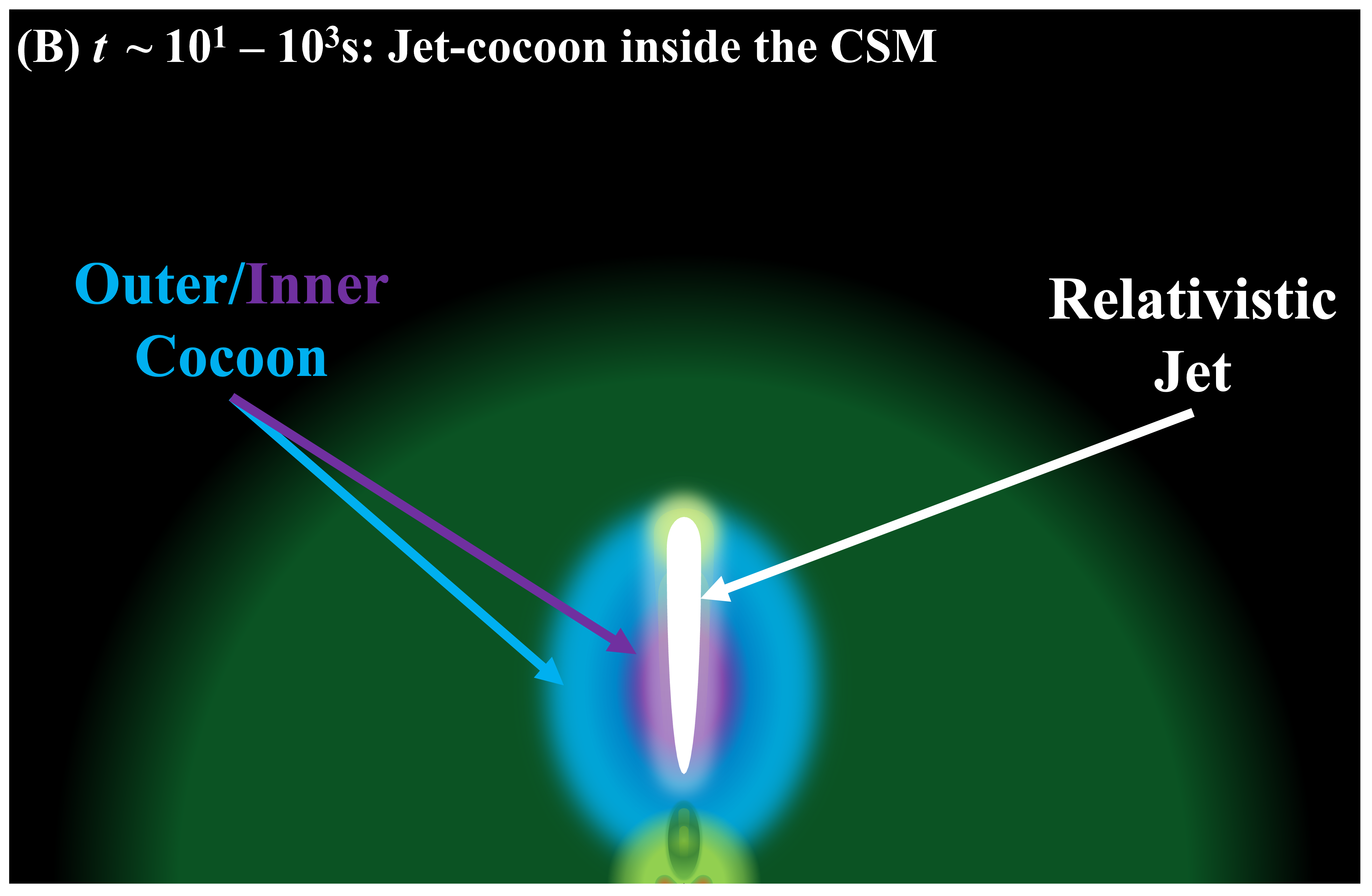} 
    \includegraphics[width=0.49\linewidth]{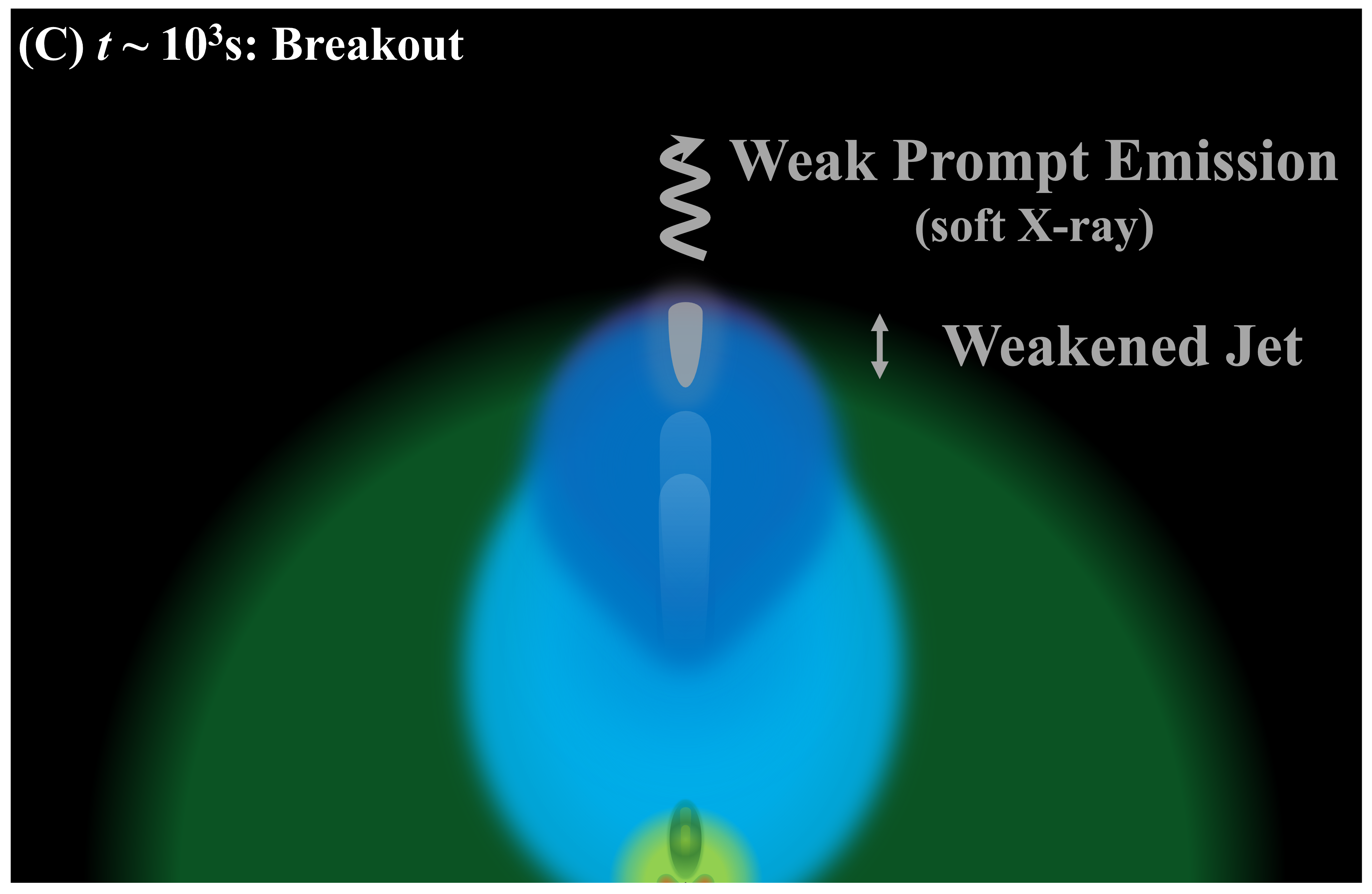} 
    \includegraphics[width=0.49\linewidth]{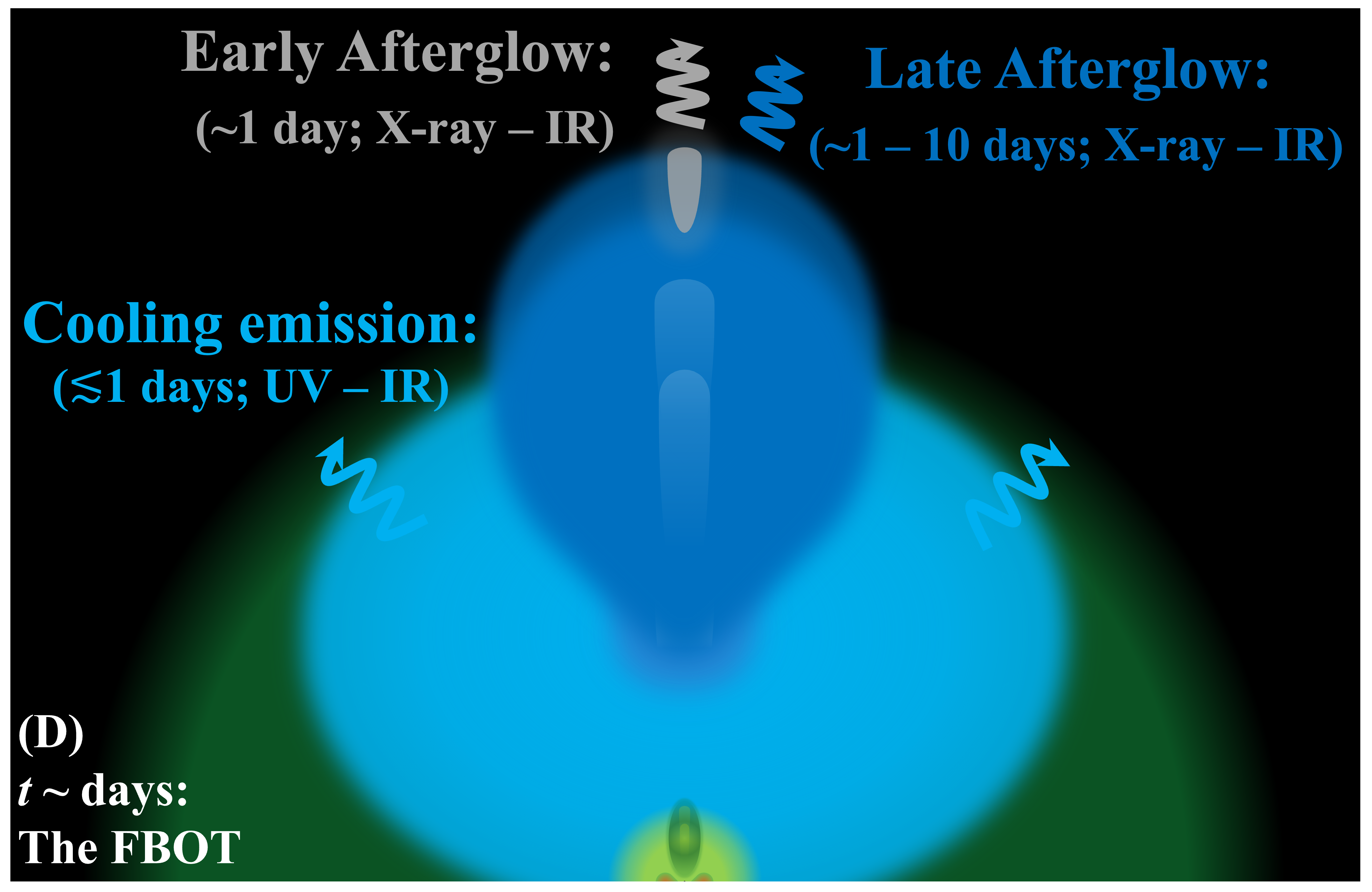} 

  \caption{Schematic picture of our scenario for EP240414a and its counterparts. 
  A) We consider an LGRB progenitor (light green) surrounded by an extended CSM (dark green) (as in \citealt{2015ApJ...807..172N}) supposedly caused by an intense mass-loss episode at the end of the progenitor star's life.  
  B) After the jet (white) breaks out of the progenitor star, it propagates in the dense CSM, forming a shock head (yellow), where the jet/CSM are shocked into an inner-cocoon (dark blue) and an outer-cocoon (light blue; respectively). 
  C) By the time the jet is about to break out of the CSM, it has been significantly weakened (gray) energetically and relativistically, while an energetic cocoon is formed.
  Consequently a weak prompt emission is produced.
  D) The sub-energetic and mildly relativistic jet and inner-cocoon power afterglow emissions, and the shocked CSM (slower outer-cocoon; light blue) powers cooling emission.
  }
  \label{fig:scheme} 
\end{figure*}

Figure \ref{fig:scheme} illustrates our scenario.
At $t_0\equiv 0$ (A, $t$ being the rest-frame time), the jet is launched from the central engine.
It quickly punches through the stellar core before entering the extended CSM (B).
As it propagates through the dense medium, the highly relativistic jet material (white) continuously mixes with the CSM at the jet head (yellow), leading to gradual energy loss and weakening of the jet.
This interaction powers a hot inner-cocoon composed of shocked jet material (mildly relativistic; see Figure 1 in \citealt{2011ApJ...740..100B}) and an outer-cocoon of shocked CSM material (non-relativistic), both of which help maintain the jet’s collimation.

By the time the jet reaches the outer edge of the CSM (C), it has been significantly weakened, with its energy reduced to $E_{\rm j} \sim \text{a few }\times10^{50}$ erg and its Lorentz factor to $\Gamma \sim 20$. Meanwhile, most of its initial energy ($E_{\rm eng} \sim 10^{52}$ erg) has been transferred to the outer-cocoon (light blue). We classify this as a ``barely failed" jet (following \citealt{2022MNRAS.517..582E}). As a result, the breakout is expected to produce a weak and soft prompt emission (C), resembling that of EP240414a.

The jet and ``inner-cocoon" remain mildly relativistic, powering luminous non-thermal afterglow emission from radio to X-ray bands (D). 
Meanwhile, the ``outer-cocoon" expands through the CSM and outward. 
As its density decreases, thermal photons escape, producing cooling emission on a timescale of $\sim 0.1 - 1$ day (in rest-frame; D). 
The combined contributions of afterglow and cooling emission result in an FBOT-like transient over the following hours to days. 
In the next sections, we explore this scenario as a potential explanation for the peculiar features of EP240414a.

We follow the same approximations as in \cite{2025arXiv250316242H} with respect to the CSM and the jet.
In particular,  we consider a conventional LGRB jet.
We consider a total energy budget of $E_{\rm{eng}}\sim 10^{52}$ erg deposited on a timescale $t_{\rm{eng}}=100$ s (which is comparable to the free-fall timescale of the WR progenitor star; \citealt{2003MNRAS.345..575M}) in the form of a relativistic jet with an initial opening angle $\theta_0=10^\circ$.
For the CSM we assume a radius $R_{\rm CSM} = 3\times 10^{13}$ cm and a mass $M_{\rm CSM} = 0.03 M_\odot$ so that the jet is barely failed (see \citealt{2025arXiv250316242H}) and thermal emission is consistent with observations (\citealt{2025ApJ...982L..47V}).

Our model is quite similar to the jet-CSM model M01R400 which has been investigated numerically in \cite{2024PASJ...76..863S}, despite their slightly larger mass $M_{\rm CSM}=0.1 M_{\odot}$ (see their Table 1) although this is expected to have a limited effect on the jet propagation (compared to $R_{\rm CSM}$; see \citealt{2025arXiv250316242H}).
Their results confirm the barely failed outcome considered here and the mildly relativistic nature of the jet (see Figure 6 in \citealt{2024PASJ...76..863S}).

\subsection{Jet's dynamics}
We solve the dynamics of jet propagation following the analytic model of \cite{2020MNRAS.491.3192H,2021MNRAS.500..627H}, based on \cite{2011ApJ...740..100B} which has been generalized in \cite{2025arXiv250316242H}, where relativistic jet head dynamics are properly taken into account.
Our analytic model solves the jet head $\beta_h$, as a function of the jet/CSM parameters, which then give the time it takes for the jet to breakout (using $t_b=R_{CSM}/(c\beta_h)$) as:
\begin{equation}
\begin{split}
t_{\rm b} \approx & 1180 \text{\,s}
\left(\frac{R_{\mathrm{CSM}}}{3\times 10^{13} \text{\,cm}}\right)^{\frac{2}{3}} 
\left(\frac{M_{\mathrm{CSM}}}{3\times 10^{-2}M_\odot}\right)^{\frac{1}{3}} 
 \\
& \times \left(\frac{\theta_{0}}{10^\circ}\right)^{\frac{2}{3}} \left(\frac{t_{\rm{eng}}}{10^2\text{\,s}}\right)^{\frac{1}{3}} \left(\frac{E_{\mathrm{eng}}}{10^{52} \text{\,erg}}\right)^{-\frac{1}{3}} 
\left(\frac{1-\beta_{\rm h}}{0.17}\right)^{-\frac{5}{3}}.
\label{eq:tb apporx}
\end{split}
\end{equation}

The general criterion for jet failure is $t_{\rm b} \geqslant {t_{\rm eng}}/({1-\beta_{\rm h}})$ (see \citealt{2025arXiv250316242H}; also see \citealt{2023MNRAS.519.1941P}).
We consider the limit $t_{\rm b}\sim  {t_{\rm eng}}/({1-\beta_{\rm h}})$, which is where the ``barely failed" jet is favored (see \citealt{2025arXiv250316242H}).
Hence, in its final form, the jet is weakened; as it has lost most of its energy to the cocoon and suffered from baryon loading.

\subsection{Post-breakout evolution}

\subsubsection{Residual mildly relativistic jet \& inner-cocoon}
\label{sec:residual jet}
In the barely failed jet regime, the degree of mixing is important but difficult to estimate analytically (\citealt{2017ApJ...834...28N,2023MNRAS.520.1111H}). However, our model closely resembles M01R400 in \cite{2024PASJ...76..863S}, whose numerical simulations confirm our analytic predictions and support the barely failed jet scenario.
Their numerical simulations show that the sub-relativistic outer-cocoon component ($\Gamma\beta\lesssim 1$) dominates in terms of energy ($E_{\rm c}\sim 10^{52}$ erg), 
and the energy of the residual jet (the jet and the inner-cocoon not well mixed with the outer-cocoon) is of the order $E_{j}\sim 10^{50}-10^{51}$ erg with $\Gamma\sim \text{a few}-\text{a few}\times10$ (see Figure 6 in \citealt{2024PASJ...76..863S}).

Furthermore, numerical simulations indicate that after breaking out, the jet develops an angular structure, with the Lorentz factor and energy density varying as a function of angle (\citealt{2021MNRAS.500.3511G}). This angular structure plays a crucial role in shaping the afterglow emission. To account for this and following the results of \cite{2024PASJ...76..863S}, we adopt a simplified model consisting of two residual jet components:
\begin{enumerate}
    \renewcommand{\labelenumi}{\roman{enumi})}
    \item Mildly relativistic core jet: with $\Gamma\sim \text{a few } \times10$, and $E_{\rm j,core}\sim 10^{50}-10^{51}$ erg (gray in Figure \ref{fig:scheme}).
    \item Mildly relativistic inner-cocoon: in the jet wings with $\Gamma\sim \text{a few}$, and $E_{\rm c,rela.}\sim 10^{50}-10^{51}$ erg (dark blue in Figure \ref{fig:scheme}).
\end{enumerate}

\subsubsection{Non-relativistic outer-cocoon}
\label{sec:NR CC}
Since the jet is barely failed, the non-relativistic outer-cocoon ($\Gamma\beta< 1$) is expected to carry most of the engine energy (see orange lines in Figure 6 in \citealt{2025arXiv250316242H}; also see Figure 6 in \citealt{2024PASJ...76..863S}), hence its importance.
As shown in Figure \ref{fig:scheme} (D), this outer-cocoon component is expected to shock the CSM and expand in a quasi-spherical geometry.
Therefore, asymptotically, the outer-cocoon is expected to have a total mass $M_{\rm c}\equiv M_{\rm CSM}$.

At the breakout time the energy of the outer-cocoon is dominated by the internal energy ($E_{\rm c,i}\sim E_{\rm c}$).
As the outer-cocoon breaks out and expands radially [to $r(t)$], it cools adiabatically, and its internal energy is converted to kinetic energy as $E_{\rm{c,i}}(t)\approx E_{\rm{c,i}}(t_b){R_{\rm CSM}}/{r(t)}$ .
Once a considerable amount of internal energy is converted to kinetic energy, the expansion can be approximated to be homologous so that ${r(t)}=R_{\rm CSM}+ {c (t-t_b)\beta(r) }$,
which further simplifies to ${r(t)}\approx {c t\beta(r) }$
for $t\gg t_{\rm b}$ and $r(t)\gg R_{\rm CSM}$.

Based on numerical simulations (e.g., \citealt{2017ApJ...834...28N,2022MNRAS.517..582E,2023MNRAS.519.1941P}) one can consider that the distribution of $\frac{dE_c}{d\log(\Gamma\beta)}\sim \text{Const}$.
Energy conservation gives the asymptotic density profile of the outer-cocoon as $\rho_c\propto \beta^{-5}$
(\citealt{2022ApJ...925..148S,2022MNRAS.517..582E}).

The typical asymptotic velocity of the outer-cocoon as $\beta_t\equiv \langle\beta_c\rangle\sim \sqrt{2E_c/(M_{\rm CSM}c^2)}$ ($\sim 0.5$ here).
Considering that $\frac{dE_c}{d\log(\Gamma\beta)}\sim \text{Const}$, and $\Gamma\beta\sim \beta$, we approximate that half of the outer-cocoon energy is in the part with $\beta>\beta_t$ and the other half is in the part with $\beta<\beta_t$.
Given the steep density profile, $M_c(>\beta)\propto\beta^{-2}$, one can infer that most of the mass is contained in the slower outer-cocoon part ($\beta_0<\beta<\beta_t$; where $\beta_0$ can be found from mass and energy conservation equations).

Here, we focus on the fast and luminous cooling emission, which manifests as an FBOT similar to the one observed in EP240414a, occurring on a timescale of hours to days.
The very small mass in the part with $\beta>\beta_t$ implies very short diffusion timescales.
The near-relativistic velocities ($\beta \gtrsim 0.5$) further shorten the observed times (with $t_{obs} = (1-\beta)t_{lab}$).
These two effects make emission from this component shine (and decay) very early (sub - hour timescales) in UV bands (see \citealt{2023MNRAS.524.4841H} for a very similar emission).
In the context of EP240414a, emission from the faster part is too early and does not coincide with any observational data in UV/Optical/IR bands.
Hence, emission from the faster outer-cocoon part will not be considered in the following, and we only consider the slower part ($\lesssim 0.5c$) which coincides with observational data available in timescales of hours -- days.

\subsection{Emission}
\subsubsection{Cooling emission from the outer-cocoon}
\label{sec:cc emi}
We estimate the cooling emission from the non-relativistic outer-cocoon (light blue in Figure \ref{fig:scheme}).
We follow the same analytic model in \cite{2024ApJ...963..137H,2024ApJ...971L..30H}.
The procedure is as follows.
First, using the homologous expansion approximation, the density profile and optical depth can be found as a function of velocity and observer time [with $t_{\rm obs}=(1+z)t$]. 
We assume a gray opacity $\kappa\sim 0.07 \text{ g$^{-1}$ cm$^2$}$ dominated by Thomson scattering (for $Z=1$; see \citealt{2019A&A...621A..71T,2025ApJ...982L..47V}).
The diffusion velocity, from where radiation is decoupled with matter, can be found using the criterion $\tau_{\rm d}\sim 1/(\beta_m-\beta_{\rm d})$ (see \citealt{2015ApJ...802..119K,2023MNRAS.524.4841H}).
Photospheric velocity can be found using $\tau_{\rm ph}\sim 1$.
This gives the time evolution of $\beta_{\rm d}(t_{\rm obs})$ and $\beta_{\rm ph}(t_{\rm obs})$.

Taking into account the adiabatic cooling, the amount of energy available to diffuse out (in the form of blackbody emission) as a function of observed time can be found analytically 
(see Appendix C in \citealt{2023MNRAS.524.4841H}).
Temperature can then be found using Stefan-Boltzmann equation at the photosphere.

Our analytic modeling for cooling emission assumes a sharp diffusion shell (\citealt{2010ApJ...725..904N,2012ApJ...747...88N}), 
and once the diffusion photosphere has reached the inner velocity at $t_{\rm obs,d}$, the entire outer-cocoon is optically thin.
To account for late emission, we assume that the late-time luminosity decays exponentially as $\propto \exp{\left[-\frac{1}{2}\left(\frac{t_{\rm obs}^2}{t_{\rm obs,d}^2}-1\right)\right]}$, following the same treatment as in (\citealt{2015ApJ...808L..51P,2021ApJ...909..209P}).

In the following, cooling emission is estimated for the non-relativistic outer-cocoon using: $\beta_0\sim 0.38$, $\beta_t\sim 0.54$, $E\sim E_{\rm eng}/2$, and $M_c\sim M_{\rm CSM}$.

\subsubsection{Afterglow emission}
\label{sec:AG emi}

As explained in Section \ref{sec:residual jet}, we consider two jet components [see (i) and (ii) above],
in which the overall afterglow emission is given by a superposition of the radiation from two top-hat jet components. 
To calculate the afterglow flux, we use the afterglow module of {\sc AMES} (Astrophysical Multimessenger Emission Simulator), following the treatments described in \cite{2021ApJ...920...55Z,2025JHEAp..45..392Z} (see also \citealt{2011ApJ...732...77M,2024PhRvD.110f1307S}).

The code assumes a top-hat jet structure with a half-opening angle, an initial Lorentz factor, and an isotropic kinetic energy.
As the jet propagates through an ambient interstellar matter (ISM) with a given density (taken as constant) , external forward shocks are formed. 
We assume a power-law electron energy distribution with a spectral index $p$, and constant microphysics parameters $\epsilon_{\rm e}$, $\epsilon_{\rm B}$, and $f_e$, which are the energy fractions of the internal energy going into non-thermal electrons, magnetic fields, and the number fraction of accelerated electrons, respectively. 
Then, the synchrotron emission is calculated numerically.
The flux is taken by integrating the emissivity over the equal arrival time surface (\citealt{2022MNRAS.517.5541T,2025JHEAp..45..392Z}).

The afterglow model is determined by fitting the multi-wavelength data (X-ray to radio) of EP240414a.
The early X-ray (phase I) and radio ($<10$ day) data are fitted with the core jet component, which is expected to peak first, while the later X-ray and radio data are fitted with a mildly relativistic inner-cocoon component.
This component is set to explain the optical/IR data peaking at $\sim 3-5$ days, as it has been discussed as having a non-thermal origin (see \citealt{2025ApJ...978L..21S}).

We consider that the jet is observed on-axis ($\theta_v=0$), which is based on the similarity of the afterglow evolution in EP240414a to that in conventional on-axis LGRBs (see Figure 2 in \citealt{2024arXiv240919055B} and Figure 3 in \citealt{2024arXiv241002315S}).
We take $p=2.4$ based on the observed X-ray photon index (\citealt{2024arXiv241002315S}).
The remaining macroscopic parameters are adjusted around the following values to match observations: $\epsilon_{\rm B}\sim 10^{-4}-10^{-2}$, $\epsilon_{\rm e}\sim 0.1$, and $f_e\sim 0.1$.
For the ISM density, we assume $n_0 \sim 0.5$ cm$^{-3}$.

\subsubsection{The SN}
\label{sec:emi SN}
For the SN component (SN~2024gsa), based on its similarity to SN~1998bw (\citealt{2024arXiv241002315S,2025ApJ...982L..47V}), and for simplicity, we use the template of SN~1998bw in \cite{2011AJ....141..163C} at $z=0.401$.
This is a crude assumption, as a detailed study of the SN is beyond the scope of this work.

From the analysis of SN~2024gsa and its comparison with other BL Ic SNe (\citealt{2019A&A...621A..71T}; also \citealt{2016MNRAS.458.2973P,2018A&A...609A.135S,2024ApJ...976...71S}), we found evidence of moderate host extinction, which is also consistent with the high column density measured in X-ray. 
Our analysis indicates $E(B-V)\sim 0.15 -0.3$.
In the following, we adopt a conservative value $E(B-V)\sim 0.15$ (see Appendix \ref{sec:color}).

\begin{figure*}
      \centering
      \includegraphics[width=0.83\linewidth]{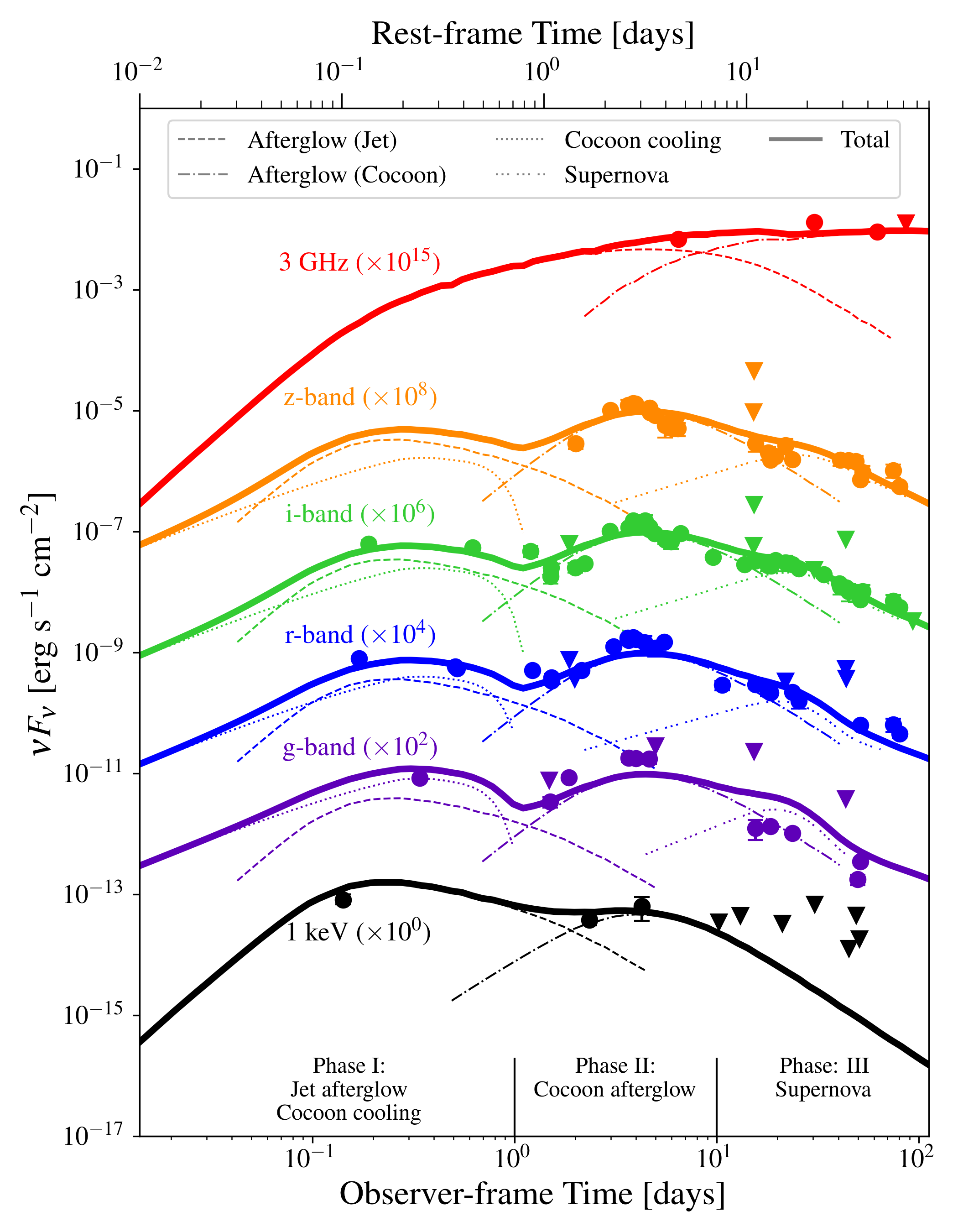} 
  \caption{Multi-band observations of EP240414a's counterparts and the theoretical interpretation.
  Observed fluxes in X-ray, g-band, r-band, i-band, z-band, and radio are shown (black, purple, blue, green, orange, and red, respectively; circles for detections and downward triangles for upper limits) in the observer-frame (bottom) and in the rest-frame (top) time (for $z=0.401$).
  Estimated total flux contributions (solid lines) along with individual components are shown: afterglow emission from the mildly relativistic jet (dashed) and the inner-cocoon (dashed dotted) [see Section \ref{sec:AG emi}]; cooling emission from the non-relativistic outer-cocoon (dotted) [see Section \ref{sec:cc emi}]; and the SN component (spaced-dotted) [see Section \ref{sec:emi SN}].
  The three phases are indicated with the corresponding dominant contributions (as in \citealt{2024arXiv241002315S}).  
  Photometric Data is taken from \cite{2024arXiv241002315S,2025ApJ...982L..47V,2025ApJ...978L..21S} and has been corrected for Galactic extinction [$E(B-V)=0.032$]. 
  In addition, moderate host extinction [$E(B-V)=0.15$] is assumed based on the analysis of the SN color and the X-ray column density (see Appendix \ref{sec:color}).
  The X-ray data are taken from Table 4 in \cite{2024arXiv241002315S}.
  Radio data are taken from Table 1 in \cite{2024arXiv240919055B}.
  A summary of the model parameters is given in Table \ref{tab:model_params} in Appendix \ref{ap:para}.
  }
  \label{fig:EP24} 
\end{figure*}


\section{Results}
\label{sec:results}
Figure \ref{fig:EP24} shows our multi-wavelength modeling of EP240414a/SN~2024gsa, in radio -- X-ray bands, in three time phases ($<1$, $1-10$, and $>10$ days; respectively; similar to \citealt{2024arXiv241002315S}).
We consider a scenario where a conventional LGRB is produced in progenitor with an extended CSM (see Figure \ref{fig:scheme}), which produces a barely failed jet -- an intermediate jet between failed (LLGRBs) and successful jets (LGRBs).
Our model consistently reproduces the observed data (with minor exceptions) in each of the three phases, and suggests that such intermediate jet offers a coherent explanation to EP240414a and its unique features in comparison to typical LGRBs/LLGRBs.

\subsection{Cooling emission (phase I)}
\label{sec:cooling emission}

The bolometric luminosity of the outer-cocoon cooling emission is found as:
\begin{align}
    L_{\rm bol}(t_{\rm obs}) \sim & \,1.7 \times 10^{46} \text{ erg s}^{-1} 
    \left(\frac{t_{\rm obs}}{0.3 \text{ day}}\right)^{-k} \notag \\
    & \times \left(\frac{E_{\rm eng}}{10^{52} \text{ erg}}\right) 
    \left(\frac{R_{\rm CSM}}{10^{13.5} \text{ cm}}\right),
    \label{eq:Lbol_approx}
\end{align}
with $t_{\rm obs,p} \sim 0.3 \text{ day} 
\left(\frac{\kappa}{0.07 \text{ cm}^2 \text{ g}^{-1}}\right)^{\frac{1}{2}} 
\left(\frac{M_{\rm CSM}}{0.03 M_\odot}\right)^{\frac{1}{2}}\left(\frac{1+z}{1.4}\right)$ is the observed peak time (in flux and magnitude), and $k$ evolving from $\sim 1.5$ ($t_{\rm obs}\ll t_{\rm obs,p}$) to $\sim 2$ ($t_{\rm obs}\sim t_{\rm obs,p}$).

This high luminosity is due to the large radius at which internal energy is deposited (breakout radius $R_{\rm CSM}\sim 10^{13.5}$ cm), which reduces internal energy loss to adiabatic cooling.
The short observed timescale makes such transients difficult to detect, if not for EP's rapid and accurate localization.

In Figure \ref{fig:EP24}, our model of cooling emission (dotted line; phase I) is consistent with optical photometric data (both flux and color) at $\sim 0.1 - 1$ day.
Our model predicts high-temperature thermal emission during this phase, which aligns with the very blue optical spectrum observed at $\sim 0.6$ days (see Section 3.3 in \citealt{2025ApJ...982L..47V}), coinciding with the decay phase of this cooling emission.
At that time, the thermal component dominates at shorter wavelengths (e.g., g-band) due to its high temperature ($\sim 3\times 10^4$ K), while at longer wavelengths (e.g., i-band), it contributes comparably to the afterglow.
This naturally explains the observed blue spectral slope and its evolution at later times (see Figure 2 in \citealt{2025ApJ...982L..47V}).
Thus, our scenario provides a consistent interpretation of both the photometric and spectroscopic data in phase I.

\cite{2025ApJ...982L..47V} conducted a detailed analysis incorporating multiple physical components to model this early phase but found that the cocoon emission model would require an extreme energy budget of $\sim 10^{53}$ erg.
Our barely failed jet scenario with an extended CSM reduces adiabatic energy losses and provides a reasonable fit (combined with jet afterglow; see Section \ref{sec:AG result}) with a lower energy requirement of $\sim 10^{52}$ erg.

In our scenario, cooling emission from the inner-cocoon is also expected, but in very short timescales ($\sim 10^2 -10^3$ s after the breakout; see \citealt{2023MNRAS.524.4841H}; and potentially in soft X-ray see \citealt{2024ApJ...963..137H}) which would coincide with the prompt emission.
This has not been estimated here because of the limited early data to compare with but future high-cadence surveys might reveal more about cooling emission from the inner-cocoon (e.g., ULTRAST and LSST).

\subsection{Afterglow emission (phases I -- III)}
\label{sec:AG result}
Figure \ref{fig:EP24} shows our afterglow model for EP240414a.
Our model parameters are as follows\footnote{Energy in the mildly relativistic jet/inner-cocoon is merely $\sim 2.5 \%$ of the original jet energy ($E_{\rm eng}=10^{52}$ erg) which reflects the weakness of the residual jet -- due to its nature as barely failed.}:
\begin{enumerate}
    \renewcommand{\labelenumi}{\roman{enumi})}
    \item \text{Mildly relativistic core jet:} Macroscopic parameters: $\Gamma= 20$, $E_{\rm j,core}= 1.5\times 10^{50}$ erg, and $\theta_j= 0.1$ rad.
    Microscopic parameters: $\epsilon_{\rm B} = 2\times 10^{-4}$, $\epsilon_{\rm e} = 3\times 10^{-2}$, and $f_e = 0.2$
    \item \text{Mildly relativistic inner-cocoon}: Macroscopic parameters: $\Gamma\sim 5$, $E_{\rm c,rela}= 10^{50}$ erg, and $\theta_c\sim 0.2$ rad. Microscopic parameters: $\epsilon_{\rm B} = 6\times 10^{-2}$, $\epsilon_{\rm e} = 0.4$, and $f_e = 0.4$.
\end{enumerate}
General parameters are: $\theta_v=0$ rad, $n_0 = 0.6$ cm$^{-3}$, and $p = 2.4$ (also see Table \ref{tab:model_params} in Appendix \ref{ap:para}).

Our model shows that, in the optical/IR bands, afterglow emission transitions from being subdominant in phase I to becoming the dominant component in phase II (the FBOT peak; AT~2024gsa). 
This is consistent with spectroscopic observations of phase I, which suggest a thermal origin (\citealt{2025ApJ...982L..47V}), as our afterglow component (non-thermal) remains subdominant. 
Furthermore, this aligns with the findings of \cite{2025ApJ...978L..21S}, which indicate that the peak of phase II is inconsistent with a thermal origin, as it would require a very large photosphere (moving with $\sim 0.7c$) to match the color. 
Overall, our model, despite its simplicity, provides a reasonable explanation for the photometric data across all bands, with the exception of the transition between phase I and phase II. 
We suspect that this is due to our reliance on only two jet components\footnote{In reality, the jet is expected to exhibit a multi-component structure (e.g., \citealt{2021MNRAS.500.3511G}), and our model lacks an intermediate component that would peak between the jet and the inner-cocoon, which would improve the fit in this transition phase.}.
Allowing for a variation in the viewing angle near the jet axis would improve the fit across the transition phase. 
In \cite{2025arXiv250324266Z} a different viewing angle is considered, and phase~II (its rise in particular) is explained by a top-hat relativistic jet viewed with $\theta_v\sim 10-15^\circ$.

The observed peaks in phases I and II correspond to the respective peaks in afterglow emission from the core jet and the inner-cocoon. 
For an on-axis jet, light curves exhibit peaks at the transition from the coasting phase to the self-similar phase (\citealt{1997ApJ...489L..37S}). 
The observer time of the flux maximum can be determined analytically as
$t_{\rm obs, peak} \sim 3(1+z)E_{\rm iso}/(32\pi n_0m_pc^5\Gamma^8)^{1/3}$,
where $E_{\rm iso}$ is the isotropic kinetic energy (\citealt{1997ApJ...489L..37S}).
For our core jet and inner-cocoon parameters, we obtain peaks at $t_{\rm obs} \sim 0.13$~days and $t_{\rm obs} \sim 3$~days, respectively, which are in good agreement with our numerical results.

Our model is also consistent with radio observations.
For the core jet, the absorption frequency $\nu_a$, the typical frequency $\nu_m$, and the cooling frequency $\nu_c$ satisfy the relation $\nu_a < \nu_m < \nu_c$ for $0.1~{\rm days}\lesssim t_{\rm obs}\lesssim 10~{\rm days}$.
The typical frequency $\nu_m$ decreases with time, and at $t_{\rm obs} \sim6$ days, it crosses the 3 GHz band.
In this regime, the radio flux follows the scaling $F_\nu \propto t_{\rm obs}^{-3(p-1)/4} \approx t_{\rm obs}^{-1}$ (\citealt{2013NewAR..57..141G}).
For the inner-cocoon, the system enters the self-similar phase after $\sim 2$ days.
For $t_{\rm obs}\gtrsim2$~days, the break frequencies $\nu_a$, $\nu_m$, and $\nu_c$ satisfy the relation $\nu_a < \nu_m < \nu_c$.
Here, $\nu_m$ remains above the radio band, while $\nu_a$ is below it.
In this regime, the radio flux follows $F_{\nu}\propto t_{\rm obs}^{1/2}$ (\citealt{2013NewAR..57..141G}), which seems to be consistent with observations.

\subsection{The Supernova (phase III)}
\label{sec:SN}
In Figure \ref{fig:EP24}, we show the template of SN~1998bw from \cite{2011AJ....141..163C} as a reference (spaced-dotted line; phase III; assumed to be powered by Ni$^{56}$ decay).
Apart from the g-band, where the observed data indicate a significantly dimmer SN, this template provides a reasonable fit.
The discrepancy in the g-band is likely due to a combination of two factors; to a lesser extent, uncertainties in the intrinsic luminosity of SN~2024gsa, which appears intermediate between SN~1998bw and SN~2006aj (\citealt{2024arXiv241002315S,2025ApJ...982L..47V}); and to a greater extent, uncertainties in the extinction at the rest-frame of EP240414a (our assumption of $E(B-V) \sim 0.15$ is conservative; see Appendix \ref{sec:color}).
Although the nature of SN~2024gsa is intriguing, it is beyond the scope of this study, which focuses on jet-driven transients evolving on timescales of $<10$ days.


\section{Discussion \& Conclusions}
\label{sec:diss}

\begin{figure}
    \centering
    \includegraphics[width=0.99\linewidth]{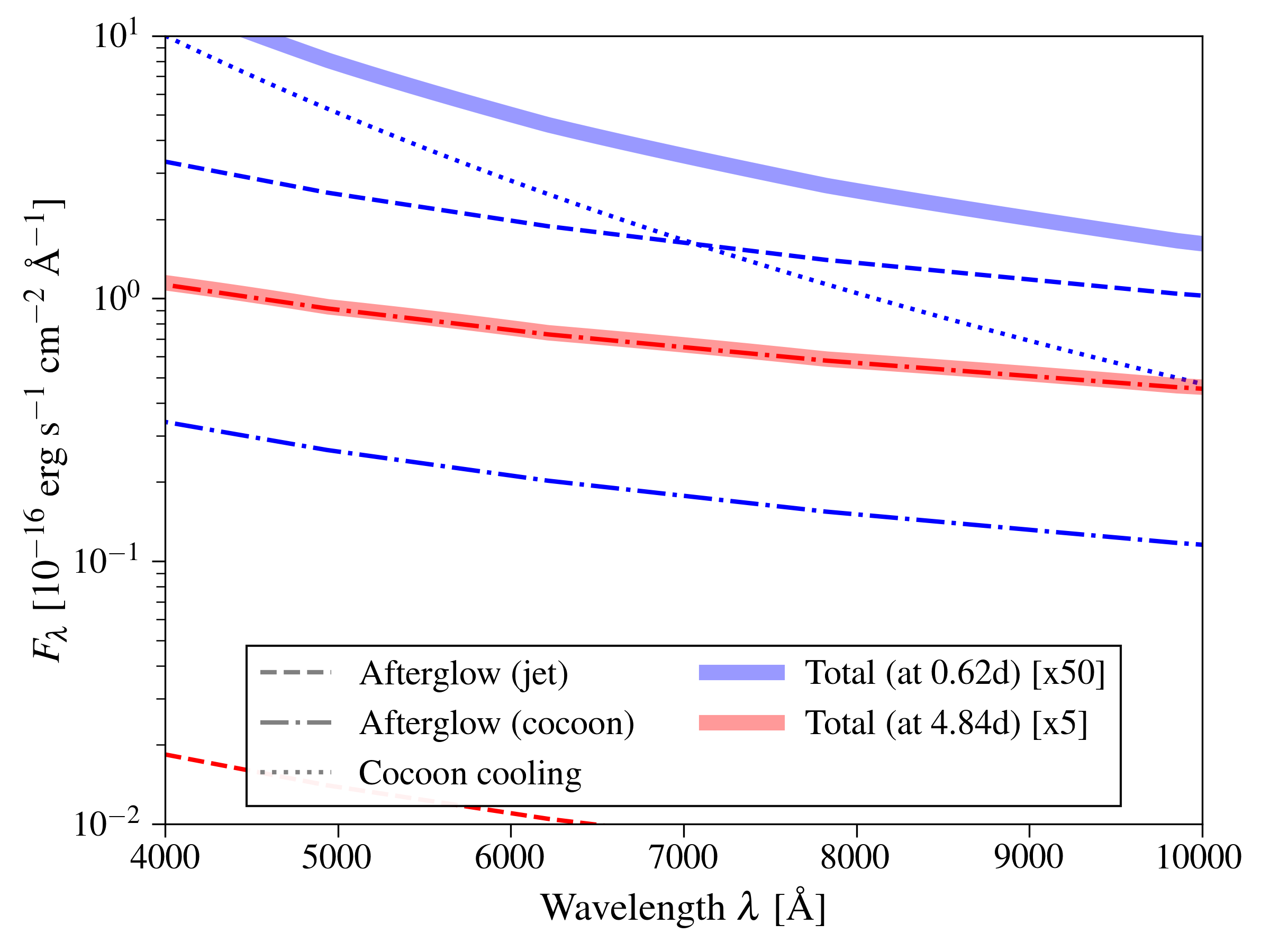} 
  \caption{Spectral energy distribution as a function of wavelength, in the optical range, showing the contributions from the jet afterglow (dashed), cocoon afterglow (dash-dotted), cocoon cooling emission (dotted), and the sum of all contributions (shaded). The spectral energy distribution is shown at 0.62 days during phase I (first blue peak; blue) and at 4.84 days during phase II (second red peak; red). No correction for Galactic or host-galaxy extinction has been applied.
  The spectral energy distribution is scaled to the observed spectra (reported in Figure~2 of \citealt{2025ApJ...982L..47V}) and shows good agreement.
  }
  \label{fig:SED} 
\end{figure}

\begin{figure}
    \centering
    \includegraphics[width=0.99\linewidth]{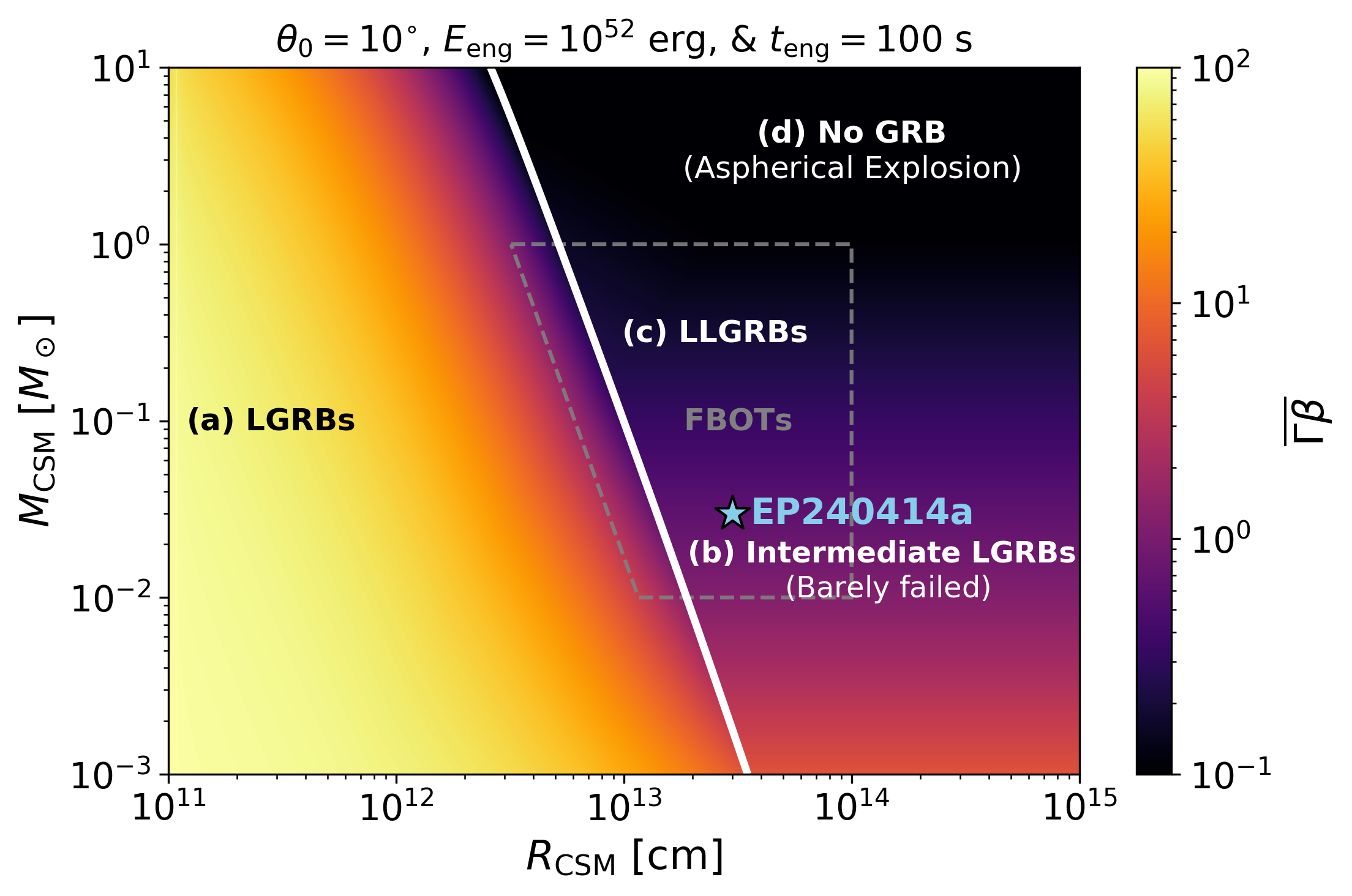} 
  \caption{The typical proper velocity of the jet-cocoon system after propagating through various CSM environments, characterized by their mass ($M_{\rm CSM}$) and radius ($R_{\rm CSM}$). 
  Different outcomes are indicated: (a) LGRBs, powered by highly relativistic jets ($\overline{\Gamma\beta}\sim 10-100$); (c) LLGRBs, originating from shock breakouts of failed jets ($\overline{\Gamma\beta}\sim 0.1$); (b) Intermediate LGRBs, arising from ``barely failed" jets ($\overline{\Gamma\beta}\sim 1$); and (d) cases where no GRB is produced, resulting only in aspherical or quasi-spherical cocoon-driven explosions ($\overline{\Gamma\beta}\lesssim 0.1$).
  The gray region indicates where luminous FBOTs from outer-cocoon cooling emission are expected.
  The star symbol marks the CSM conditions adopted here for EP240414a (as modeled in Figure~\ref{fig:EP24}), emphasizing its intermediate nature between LGRBs and LLGRBs.
  A conventional LGRB jet is assumed (see title).
  For further details, refer to \cite{2025arXiv250316242H}.}
  \label{fig:phase} 
\end{figure}

We investigated the nature of EP240414a which exhibits several unusual features: (1) an unusually dim and soft LGRB-like prompt emission, (2) a luminous FBOT (AT~2024gsa) with (3) an unusually red color at its peak, and (4) a BL Type Ic SN (SN~2024gsa) [see Section \ref{sec:2}]. 
These observational properties suggest that EP240414a represents an intermediate case between conventional LGRBs and LLGRBs.

To explain these features, we proposed a ``barely failed jet" scenario, where a conventional LGRB jet ($\sim 10^{52}$ erg) propagates through a collapsar progenitor surrounded by an extended CSM ($R_{\rm CSM} \sim 3\times 10^{13}$ cm, $M_{\rm CSM} \sim 0.03 M_\odot$; see Figure \ref{fig:scheme}). 
Our analytic modeling of jet propagation in \cite{2025arXiv250316242H}, supported by numerical simulations (M01R400 in \citealt{2024PASJ...76..863S}), indicates that the jet is significantly weakened before the breakout, producing a mildly relativistic residual jet (with $\sim 10^{50}-10^{51}$ erg, $\Gamma \sim 20$). 
Such an intermediate state could explain the intermediate nature of the prompt emission of EP240414a.

As shown in Figure \ref{fig:EP24}, our model successfully reproduces the observed multi-wavelength light curves. 
We found that thermal cooling emission from the non-relativistic outer-cocoon dominates in phase I ($<1$ day), while non-thermal afterglow emission from the mildly relativistic jet-cocoon explains the peak of AT~2024gsa (phase II; $1-10$ days). 
This is consistent with spectroscopic analyses indicating early thermal emission (in phase I; see \citealt{2025ApJ...982L..47V}) and a non-thermal origin for the FBOT peak (phase II; see  \citealt{2024arXiv241002315S,2025ApJ...982L..47V,2025ApJ...978L..21S}).

In Figure~\ref{fig:SED}, we show the spectral energy distribution of our model, including contributions from the afterglow (jet and cocoon; non-thermal) and cocoon cooling emission (thermal), at the first peak (0.62 days; phase~I) and the second peak (4.84 days; phase~II). Comparison with the observed spectra from Figure~2 of \cite{2025ApJ...982L..47V} shows good agreement between our model and observations at both epochs.\footnote{The spectral energy distribution shown in Figure~\ref{fig:SED} has not been corrected for Galactic or host-galaxy extinction (see Appendix~\ref{sec:color}). Accounting for extinction would slightly reduce the flux and further improve consistency with the observations.} This supports the validity of our model and, in particular, demonstrates that the rapid color evolution from blue to red is driven by the short-lived cocoon cooling emission, which dominates only during phase~I.

We adopt an on-axis viewing angle in our model ($\theta_{\rm v} \sim 0^{\circ}$), motivated by the very bright optical emission (in phases~I and II) and its likely afterglow origin (see Figure \ref{fig:SED}, and Figure 2 in \citealt{2025ApJ...982L..47V}).
Our scenario is particularly appealing as the required kinetic energy of the jet in the afterglow phase is of similar order to the energy of the prompt emission ($\sim 10^{50}$ erg; \citealt{2024arXiv241002315S}).
A slightly off-axis viewing angle remains possible, provided that the observer's viewing angle is not significantly larger than the jet-cocoon opening angle ($\theta_{\rm v} \lesssim \theta_{\rm j}$).
A significantly off-axis configuration ($\theta_{\rm v} \gg \theta_{\rm j}$) would be difficult to reconcile with the observations in our scenario, as it makes the prompt emission and the early afterglow emission (in optical) much dimmer.
Different configurations with off-axis jets  may also be possible (see \citealt{2025arXiv250324266Z}).

Our barely failed jet scenario naturally explains the intermediate nature of EP240414a with an intermediate jet (relativistically and energetically), linking it to an intermediate state between complete jet failure ($\Gamma\beta\sim 0.1$; producing LLGRBs \citealt{2012ApJ...747...88N,2015ApJ...807..172N}) and the state of complete jet success ($\Gamma\beta\sim 100$; producing conventional LGRBs; e.g., \citealt{2011ApJ...739L..55B}).

In Figure \ref{fig:phase}, we show a map illustrating the impact of diverse CSM environments around the LGRB progenitor on the relativistic nature of the system (i.e., the typical proper velocity of the system $\overline{\Gamma\beta}$; see \citealt{2025arXiv250316242H}).
The expected parameter space for successful/failed jets and LGRBs/LLGRBs (right/left to the thick line; $\overline{\Gamma\beta}\sim 10-100$/$\overline{\Gamma\beta}\sim 0.1$; respectively) is indicated.
The parameter space for barely failed jets is also shown ($\overline{\Gamma\beta}\sim 1$).
We show that the CSM parameters used here to explain EP240414a/AT~2024gsa (star symbol) lie at the boundary between successful and failed jets, placing them within the barely failed jet parameter space. This strongly supports our proposed barely failed jet scenario for EP240414a/AT~2024gsa.

This illustrates that EP240414a might be physically distinct from both (a) conventional LGRBs and (c) LLGRBs, providing a possible explanation for its outlier status in the Amati-Yonetoku relations (\citealt{2002A&A...390...81A,2004ApJ...609..935Y}; as these relations may arise from viewing angle effects in LGRB jets; see \citealt{2019NatCo..10.1504I}).

Comparing EP240414a to LLGRBs gives a distinct difference in the form of a strong afterglow component that powers the second optical peak (phase~II) \citep{2024arXiv241002315S,2025ApJ...982L..47V,2025ApJ...978L..21S}. 
However, if EP240414a were viewed off-axis, at a large viewing angle ($\theta_v \gg 1/\Gamma$, with $\Gamma \sim 5$ here), the afterglow-powered optical component (phase~II) would be beamed out, and the light curve would evolve from three peaks (phases~I to III) into two thermal peaks (phases~I and III, originating from cocoon cooling and the supernova, respectively), which would appear as similar to GRB~060218 \citep{2006Natur.442.1008C,2015ApJ...807..172N}.

Given the apparent observational similarity between a completely failed jet (as in GRB~060218) and a barely failed jet viewed off-axis (as in our scenario for EP240414a), despite their fundamentally different physical origins, the question is: how can one distinguish between the two scenarios?
A promising diagnostic is radio observation. In our scenario for EP240414a, the off-axis jet-cocoon radio afterglow is expected to exhibit a rising phase followed by a decline at late times (after $\sim 10$--$100$ days, as determined by the transition time to the mildly-relativistic quasi-spherical state). 
In contrast, for a completely failed jet such as in GRB~060218, the radio afterglow is expected to be powered by mildly-relativistic, quasi-spherical ejecta, resulting in a steadily declining radio emission (see Figure~2 in \citealt{2006Natur.442.1014S}).
Hence, while the optical light curve of an off-axis EP240414a may resemble that of GRB~060218, the radio afterglow provides a key diagnostic to distinguish between a barely failed jet viewed off-axis and a fully failed jet.
This highlights the importance of late-time radio observations.

Another EP event of interest is EP250108a, which appears broadly consistent with GRB~060218 \citep{2025arXiv250408886E,2025arXiv250408889R,2025arXiv250417516S,2025arXiv250417034L}, although no radio counterpart has yet been detected.
Much later-time radio observations will be required to constrain the jet-cocoon component (see Figure~10 in \citealt{2025arXiv250408886E}; Figure~7 in \citealt{2025arXiv250417516S}; and Extended Figure~4 in \citealt{2025arXiv250417034L}).

\cite{2024arXiv241002315S} estimate the lower limit of the event rate of EP240414a to be $\sim 0.3\text{ Gpc$^{-3}$ yr$^{-1}$}$. 
For comparison, LGRBs have an estimated event rate of $\sim 1\text{ Gpc$^{-3}$ yr$^{-1}$}$, while LLGRBs have a significantly higher event rate of $\sim 100\text{ Gpc$^{-3}$ yr$^{-1}$}$ 
(see \citealt{2005MNRAS.360L..77C}; \citealt{2006Natur.442.1011P}; \citealt{2006Natur.442.1014S}; 
\citealt{2006ApJ...645L.113C}; \citealt{2007ApJ...662.1111L}; \citealt{2007ApJ...657L..73G}). 
Although the event rate estimate is highly uncertain (due to observational limitations and intrinsic jet opening angle uncertainties), it may suggest that barely failed jets occur at a rate comparable to LGRBs, making them significantly rarer than LLGRBs.
This, in turn, may indicate that totally failed jets are more commonly produced, likely due to weaker jet launching conditions, or alternatively, due to denser CSM environments resulting from significant mass loss.

Recent observations of stripped-envelope CCSNe (Types Ib, Ic, Ibn, and Icn) show mounting evidence for the presence of CSM (e.g. \citealt{2007Natur.447..829P,2007ApJ...657L.105F,2008MNRAS.389..131P,2008ApJ...674L..85I,2006ApJ...653L.129B,2020A&A...643A..79S,2020MNRAS.492.2208C,2024ApJ...977..254D,2024ApJ...977....2P}).
This suggests that CSM environments are more common around stripped massive stars (WR, stars similar to LGRB progenitors) than previously thought, although LGRB jets seem to be surrounded by less impactful CSM environments (see Figure \ref{fig:phase}).

From the dynamics of relativistic jets (\citealt{2003MNRAS.345..575M,2011ApJ...740..100B,2015ApJ...807..172N}; \citealt{2025arXiv250316242H}), it is understood that CSM environments have observationally measurable consequences on collapsar jets and their transients, allowing us to infer CSM properties and the intrinsic energetics of GRBs.
Our study underscores the role of different jet components, ranging from mildly relativistic jets to non-relativistic cocoons; in shaping EP240414a and other GRB transients.
We predict that more LGRBs that exhibit CSM interaction signatures will be discovered, particularly through blue thermal cocoon cooling emission on timescales of days, with future high cadence surveys (e.g., LSST, ULTRAST).
A notable example is EP250108a, identified at the time of writing, 
which appears consistent with an FBOT powered by the cooling emission of a completely failed jet (\citealt{2025arXiv250408886E,2025arXiv250408889R,2025arXiv250417516S,2025arXiv250417034L}).
Additionally, our results highlight how afterglow emission from a mildly relativistic LGRB jet can mimic (and may account for some) luminous FBOTs.
Early multi-wavelength photometric and spectroscopic follow-up of GRBs will refine our understanding of jet-driven components and provide deeper insights into the diversity of their progenitors.


\section{Data Availability}
The code used in this article is publicly available via Zenodo: \dataset[10.5281/zenodo.15349354]{https://doi.org/10.5281/zenodo.15349354}. 
Additional data underlying this article are available from the corresponding author upon reasonable request.

\begin{acknowledgements}
We thank
Katsuaki Asano,  
Nanae Domoto,  
Sho Fujibayashi,  
Kenta Hotokezaka,  
Susumu Inoue,  
Christopher Irwin,
Wataru Ishizaki,  
Tomohisa Kawashima,
Shota Kisaka,  
Tatsuya Matsumoto,  
Takashi Nagao,
Ehud Nakar,  
Hirofumi Noda,
Ryo Sawada,  
Jiro Shimoda,
Akihiro Suzuki,
Kenji Toma,
Seiji Toshikage,
and Ryo Yamazaki
for their fruitful discussions and comments. 

This research was supported by the Japan Society for the Promotion of Science (JSPS) Grant-in-Aid for Scientific Research No. 23K19059, 24K00668, 23H04899 (K. K.),
No. 25KJ0010 (Y.S.),
and No. 22H00130, 23H05430, 23H04900, 23H01172 (K.I.).

Numerical computations were achieved thanks to the following: Cray XC50 of the Center for Computational Astrophysics at the National Astronomical Observatory of Japan, and Cray XC40 at the Yukawa Institute Computer Facility.
\end{acknowledgements}

\bibliography{0-P9}
\bibliographystyle{aasjournal}


\appendix

\section{Extinction correction}
\label{sec:color}

Observational properties [2) and 4) in Section \ref{sec:2})] indicate a red color of the optical/IR emission.
In order to understand the origin of this color, we first examine the color evolution of the SN component (SN 2024gsa).

Spectroscopy and photometry indicate that SN~2024gsa is similar to SN~1998bw (\citealt{2024arXiv241002315S,2025ApJ...982L..47V}; although fainter).
We estimated the color evolution of SN~2024gsa (in its rest-frame, taking into account the redshift effect and the Galactic extinction $E(B-V)=0.032$; \citealt{2024arXiv241002315S,2024ApJ...977..254D,2025ApJ...978L..21S}).
We compared its color evolution to that of SN1998bw (\citealt{2011AJ....141..163C}), and to the population of BL Ic SNe (in \citealt{2019A&A...621A..71T}). 
Comparison indicates that SN~2024gsa (as observed) has a much redder rest-frame color.
Comparison with the population of BL Ic SNe in \cite{2019A&A...621A..71T} shows similar $g'-r'$ color evolution around the peak, but with a larger offset of  $\sim 0.6-0.8$ mag toward redder colors.
We deduce that this could be due to moderate extinction at the host galaxy (see \citealt{2016MNRAS.458.2973P,2018A&A...609A.135S,2019A&A...621A..71T}; also see \citealt{2024ApJ...976...71S}).

This deduction is consistent with X-ray observations, showing a high column density of $\sim 2.4^{+2.2}_{-2.0}\times 10^{21}$~cm$^{-2}$ (Table 4 in \citealt{2024arXiv241002315S} and private communication; although their values indicate large uncertainty) which exceeds the estimated Galactic column density ($\sim 3\times 10^{20}$~cm$^{-2}$).
For a typical dust-to-gas ratio,
$N_{\rm H}$ is correlated with $A_{\rm V}$ as follows $N_{\rm H} (\text{cm}^{-2}) = (2.21 \pm 0.09) \times 10^{21} A_{\rm V}(\text{mag}$) (\citealt{2009MNRAS.400.2050G}), suggesting an extinction of up to $A_{\rm V}$ $\sim 1$ mag [i.e., $E(B-V)\sim 0.3$ for $R_{\rm V}\sim 3.1$].
Hence, from the color evolution of the SN and the measured X-ray column density, we deduce moderate extinction at the rest-frame of EP240414a.

We conservatively adopt a host extinction of $E(B-V) \sim 0.15$, as a higher value would require more extreme energetics for AT~2024gsa/SN~2024gsa.
We adopt \cite{1989ApJ...345..245C} extinction law for both Galactic and host extinction.
It should be noted that even such moderate extinction combined with the redshift $z=0.401$ significantly affects the blackbody color because the observed optical emission is emitted in the UV range in the rest-frame, which has a strong dependence on the extinction model ($R_{\rm U}\sim 5$; \citealt{1998ApJ...500..525S}).

\section{Summary of Parameters}
\label{ap:para}

\begin{table}[htbp]
\centering
\caption{Emission models and their parameters.}
\begin{tabular}{lccc}
\hline
\hline
\textbf{} & \textbf{} & \textbf{Inner-} & \textbf{Outer-} \\
\textbf{Parameters} & \textbf{Jet} & \textbf{Cocoon} & \textbf{Cocoon} \\
\hline
Emission & Afterglow & Afterglow & Cooling \\
\hline
Viewing angle [$\theta_{\rm v}$, rad] & \multicolumn{3}{c}{On-axis ($=0$)} \\
\hline
Energy [erg] & $1.5\times10^{50}$ & $10^{50}$ & $5\times10^{51}$ \\
Velocity [$\Gamma\beta$] & 20 & 5 & $0.4 - 0.6$ \\
Opening angle [rad] & 0.1 & 0.2 & $\pi/2$ \\
\hline
Opacity $\kappa$ [cm$^2$ g$^{-1}$] & -- & -- & 0.07 \\
ISM density $n_0$ [cm$^{-3}$] & 0.6 & 0.6 & -- \\
Spectral index $p$ & 2.4 & 2.4 & -- \\
Electron fraction $f_e$ & 0.2 & 0.4 & -- \\
Electron energy fraction $\epsilon_e$ & $3\times10^{-2}$ & $4\times10^{-1}$ & -- \\
Magnetic energy fraction $\epsilon_B$ & $2\times10^{-4}$ & $6\times10^{-2}$ & -- \\
\hline
\hline
\end{tabular}
\label{tab:model_params}
\end{table}

\label{lastpage}

\listofchanges

\end{document}